\documentclass{aa}
\usepackage{graphicx}
\usepackage{txfonts}
\usepackage{epsfig,subfigure}
\begin{document}
   \title{Relativistic jet motion in the core of the radio-loud quasar \object{J1101+7225}}

   \subtitle{}

   \author{J.-U.~Pott \inst{1,2} \and A.~Eckart \inst{1} \and M.~Krips \inst{1
          \and 3} \and T.P.~Krichbaum \inst{4} \and S.~Britzen \inst{4} \and
          W.~Alef \inst{4} \and J.A.~Zensus \inst{4}
          }

   \offprints{J.-U. Pott, \email{pott@ph1.uni-koeln.de}}

   \institute{I. Physikalisches Institut, University of Cologne,
Z\"ulpicher Strasse 77, 50939 K\"oln,
Germany \\\email{pott;eckart;krips@ph1.uni-koeln.de}\and European Southern Observatory, Karl-Schwarzschild-Str. 2, 85748 Garching b. M\"unchen, Germany \and Institut de Radio-Astronomie
Millim\'etrique (IRAM), 300 rue de la piscine, 38406
Saint Martin d'H\`eres, France \and Max-Planck-Institut f\"ur Radioastronomie,
Auf dem H\"ugel 69, D-53121 Bonn, Germany \\ \email{tkrichbaum;sbritzen;walef;azensus@mpifr-bonn.mpg.de}}

 \date{}

   \abstract{Multi-epoch GHz Very Long Baseline Interferometry (VLBI) data of the radio-loud quasar
   \object{J1101+7225} were analyzed to estimate the proper motion of extended optically
   thin jet components. 
   Two components separated from the core could be mapped at 1.66~GHz,
   which is consistent with earlier observations. In one case we
   found evidence of high apparent superluminal motion ($\beta
   _{\rm app}=\,22.5\pm4$) at large (deprojected) distances to the core
   $(22\,{\rm mas}\sim\,4\,{\rm kpc,\,at\,}z=\,1.46
   )$. Typically in other quasars such high separation velocities are only found much
   closer to the core component.
Furthermore
   the Doppler factor, the magnetic field strength, and
   the angular size of the
   optically thick core were derived using published X-ray data.
Analysis of 5~GHz VLBI data reveals the
   existence of further jet components within the central 5~mas. Additionally the data published so far on the GHz-spectrum were discussed at all
   angular resolutions. \object{J1101+7225} turns out to be a standard quasar for studying
   different aspects of radio jet kinematics out to kpc-scales.
   
   }

   \maketitle
%

\section{Introduction}

\begin{table}
\begin{center}
 \begin{tabular}{llc}
\hline\hline
property & value & ref. \\
    \hline 
\multicolumn{1}{l}{IAU $^{\mathrm{a}}$ / other name:} & \multicolumn{2}{l}{\object{J1101+7225} / {[}HB89{]} 1058+726 } \\
     R.A. (J2000) & 11h 01m $(48.8054\pm0.0001)$s & B02 \\
     Decl. (J2000) & +72$^\circ$25'$(37.1183\pm0.6\cdot 10^{-3})$'' & B02 \\
    
     redshift & 1.46 & (b)  \\
     abs. magnitude & -27.3 mag\( \approx 10^{46.4}\,{\rm erg~s^{-1}} \) & (c) \\
     1.4~GHz radio cont & $(1451\pm30)$~mJy & (d)  \\
    \hline
 \end{tabular}

 \caption{\label{tab_general}General properties of \object{J1101+7225}. $^{\mathrm{a}}$: International Astronomical Union; (b): Jackson \& Browne~(1991); (c): V{\' e}ron-Cetty \& V{\' e}ron~(2001); (d): White \& Becker~(1992)}
\end{center}
\end{table}
The radio-loud quasar \object{J1101+7225} is a source of the Very Long Baseline Array (VLBA)
  Calibrator Survey VCS1 for phase-referencing observations (Beasley et
al. 2002, in the following B02). We observed this source within larger
  experiments, that were focused on imaging faint Seyfert galaxies. It
  turned out that \object{J1101+7225} shows a complex source structure, which
  makes it physically interesting but, on the other hand, also less suitable as
  a calibrator. Since so far very little information is available for
  individual sources of the VCS1, we present our results in this article.
Analysis of the (relativistic) kinematics of non-thermal radio sources
gives important physical insight into the inmost
regions of an active galactic nucleus (AGN). 
The
ejection of radio jet components and the jet kinematics are most probably related to the process of accretion
onto the nucleus itself. 
Quantitative estimates of physical properties, such as magnetic fields and
source sizes, can
be derived, too.
Our analysis is based on comparison of our 1.66~GHz Very Long Baseline Interferometry (VLBI) map obtained in 2002 with the
  available maps of past observations (Thakkar et al. 1995 (T95) at 1.66~GHz and
  B02 at 2.3~GHz). Further data from a multi-epoch study at 5~GHz of the
  central 5~mas of \object{J1101+7225} by Britzen (2002) and Britzen et al. (in prep.)  were used to determine the source
  structure of the central region, unresolved at lower
  observing frequencies.  

In Table \ref{tab_general} the general properties of \object{J1101+7225} are
summarized to introduce the quasar. B02 estimated the position of \object{J1101+7225} with (sub)mas-accuracy
(Table \ref{tab_general}), which is necessary for high-sensitivity phase-referencing experiments.
In
Sect.~\ref{sec_obs} details are given concerning the VLBI observations
that are the basis for this article. Flux densities, the
resulting maps, and a kinematical analysis of the mapped structure are presented in Sect.~\ref{sec_props}. Finally the results
are discussed and summarized in Sect.~\ref{sec_sum}.

\section{\label{sec_obs}Observations}

\begin{table}
{\centering \begin{tabular*}{\columnwidth}{ccc}
\hline\hline 
date of observation&obs. frequency&bit rate \& type of samp.\\ 
13 Feb 2002 & 1630.49MHz & 256~Mbps \& 2 bit \\ 
\hline
No. \& bandwidth of IF & polarization & system and correlator: \\
4 DSB-IFs \& 16~MHz & LCP & MKIV at MPIfR, Bonn \\
\hline
\end{tabular*}}

\caption{\label{tab_el028}Details of the observing schedule of the EVN
 observation with eight antennae of \object{J1101+7225}.}
\end{table}

The quasar \object{J1101+7225} was observed as a calibrator source within
    a larger European VLBI Network (EVN) experiment. The observations
    were conducted during the period 13-14 Feb 2002,  21h-09h UT,
    while the scans of \object{J1101+7225} typically lasted for three minutes at a total observing time of 1.6~h.
     We observed at 18~cm with the 100m antenna of the Max-Planck Institut f\"ur Radioastronomie (MPIfR) at Effelsberg, Germany, the 76m antenna at Jodrell Bank, UK, the 32m antenna at Medicina,
Italy, the 25m antenna at Onsala, Sweden, the 25m antennae at Shanghai and
    Urumqi, China, the 32m antenna at Torun, Poland, and the 14x25m antenna array at
    Westerbork, the Netherlands. Details concerning the observing mode are given in
    Table~\ref{tab_el028}. 
After the observations, the data were correlated at the VLBI correlator of the
    MPIfR in Bonn, Germany, and imported into the Astronomical Image Processing System (AIPS) via MK4IN (Alef \& Graham
    2002). The data were fringe-fitted and calibrated in a standard manner
    with the AIPS package and imaged using the DIFMAP VLBI package (Pearson et
    al. 1994). 

Further we present 5~GHz VLBI maps of \object{J1101+7225},
obtained in VLBA and global VLBI observations as part of a multi-epoch
VLBI study (the Caltech-Jodrell Bank flat-spectrum sample CJF; Taylor
et al. 1996; Britzen 2002). These observations aim at a statistical
investigation of the kinematics of a complete  sample of AGN (Britzen 2002; Britzen et al. in prep.). The sources were observed in
5.5 minute snapshot observations to determine the position and motion of jet components in the central
5~mas. The data were recorded over 32~MHz total bandwidth broken up into 4
baseband channels with 1-bit sampling. The recorded data were correlated in
Socorro, USA.

In Sect.~\ref{sec_propermot} we include the VLBI-maps of \object{J1101+7225} published earlier to analyze the evolution of the
structure found. T95 observed the quasar with global VLBI at 1.66~GHz in the
    CJ1 survey which later on became a part of the CJF sample. The other map at
    2.3~GHz was reduced from VLBA data (B02) via automatic imaging using the
    Caltech DIFMAP package.


\section{\label{sec_props}Radio properties of the quasar \object{J1101+7225}}

\subsection{\label{sec_flux}Flux densities}

\begin{table}
{\centering \begin{tabular}{cccc}
\hline \hline
 obs.type & freq. [GHz] & flux dens. [mJy] & \\
\hline 
 single-dish & 0.038 & \( (1.41\cdot10^4\pm 800) \) & (a) \\
 single-dish & 0.178 & \( (3900\pm 600) \) & (b) \\
 single-dish & 1.4 & \( (1451\pm 30) \) & (c) \\
 single-dish & 2.7 & \( (1070\pm 35) \) & (d) \\
 single-dish & 5   & \( (858\pm 76) \) & (e) \\
 single-dish & 22  & \( (820\pm 100)\) & (f) \\
\hline 
 local interf. & 1.4 & $(748\pm 40)$/beam & (g), res$\sim$1.5''\\
 local interf. & 8.4 & $(349 \pm 20)$/beam & (h), res$\sim$0.2'' \\
\hline 
 VLBI & 1.66 & (396\( \pm 20)\) & EVN Feb 2002\\
 VLBI & 2.3  & $(520\pm25)$/beam & B02,res$\sim$3~mas\\
 VLBI & 5  & ($139\pm7)$ & - 1991.4 - \\
 VLBI & 5  & $(282\pm14)$ & - 1993.4 - \\
 VLBI & 5  & $(337\pm17)$ & - 1996.6 - \\
 VLBI & 5  & $(439\pm22)$ & - 1999.9 - \\
 VLBI & 8.4  & $(383\pm20)$/beam & B02,res$\sim$1~mas\\
\hline
\end{tabular}}

\caption{\label{tab_ghzspec} 
Measured radio flux densities or peak brightnesses of \object{J1101+7225},
obtained with single-dish telescopes, local interferometers, and VLBI. The VLBI flux densities are our fitted core flux densities as presented in Tables~\ref{tab_gaussmod} \& \ref{tab_britzfits}
\newline 
{\em References and comments:} (a) Rees (1990), catalogue revised by Hales et. al (1995); (b) Gower et al. (1967); (c) White \& Becker(1992); (d) K{\"u}hr et al.~(1981); (e) Gregory et al.~(1996); (f) Ter{\" a}sranta et al.~(2001); (g) Xu et al.(1995) gave this peak brightness of a map, which was convolved to a circular beam of 1.5'' FWHM; (h) Patnaik et al.~(1992) found this peak brightness to be 80\% of the total flux density of 436~mJy;
\newline
The  values are plotted over frequency in Fig.~\ref{fig_ghzflux}. 
For the interferometric observations 5\% flux density errors are given.}
\end{table}
\begin{figure}

{\centering
\includegraphics[width=0.68\columnwidth,angle=-90]{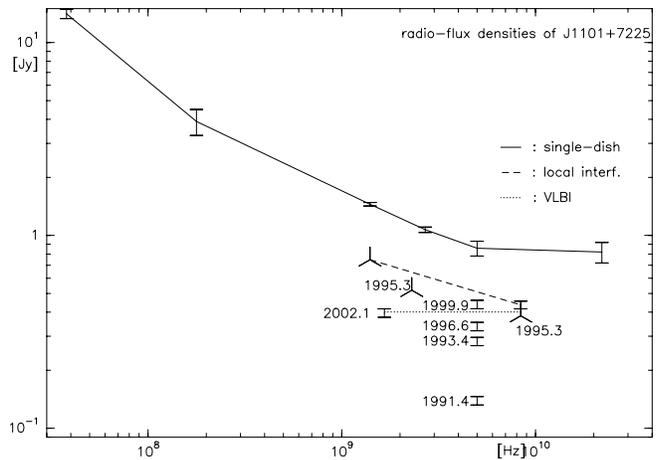}
}
\caption{\label{fig_ghzflux}The radio spectrum of
  \object{J1101+7225} at different angular resolutions, corresponding to
  Table~\ref{tab_ghzspec}. All VLBI measurements are labelled with the observing date. While the flux densities are plotted with error bars, the stars indicate peak brightnesses. Because the peak brightnesses may underestimate the flux densities, the corresponding spectra are dotted / dashed (see Sect.~\ref{sec_flux}). }
\end{figure}

\begin{table}
\begin{center}
\begin{tabular*}{\columnwidth}{cccc}
    \hline \hline
    \( S_{\rm 5~GHz}[{\rm mJy}] \)& date of observation & ref. &  ann.:\\
    \hline  
     778\( \pm 4 \) & between Feb 1977 \& Mar 1978 &  (a) &  (1)\\
     788\( \pm 31 \) &  Nov 1986 & (b) & (2)\\
     953\( \pm 73 \) & Oct 1987  & (c) & (2)\\
     623\( \pm 56 \) & 1992.5 = Jul 1992 &(d)& (3)\\
    \hline
\end{tabular*}

\caption{\label{tab_fluxvar} The published single-dish flux densities at 5~GHz
    to investigate the source variability.
(1) observed at 4.9~GHz and corrected to 5~GHz via a
spectral index \protect\( \alpha ^{\rm 10.7~GHz}_{\rm 2.7~GHz}\protect \);
(2) at 4.85~{\rm GHz}; (3) at 4.75~{\rm GHz}; (a): K{\"u}hr et al.~(1981); (b): Gregory et al.~1996); (c): Gregory \& Condon~(1991); (d): Reich et al.(2000)}
\end{center}
\end{table}


The flux density of \object{J1101+7225} has been measured occasionally
  during the last two decades.
In Table~\ref{tab_ghzspec} we summarize the measurements at the various
  frequencies. Based on these data we calculated the spectrum, shown in
  Fig.~\ref{fig_ghzflux}. 
The single-dish observations show a curved radio spectrum with a steep spectrum at low MHz frequencies ($\alpha_{\rm single}^{38{\rm M};178{\rm M}}\sim -0.8;\,S\sim \nu^{\alpha}$) and considerable flattening towards higher frequencies ($\alpha_{\rm single}^{1.4{\rm G};5{\rm G}}\sim -0.4$; $\alpha_{\rm single}^{5{\rm G};22{\rm G}}\sim 0$).
The spectral trend of the Very Large Array (VLA) measurements follows the single dish measurements. The slightly flatter shape ($\alpha_{\rm locIF}^{1.4{\rm G};8.4{\rm G}}\sim -0.3$) plotted in Fig.~\ref{fig_ghzflux} may overestimate the real spectrum, because only the peak brightness is published at 1.4~GHz. 

In contrast with the EVN we measured a core flux density of $(396 \pm 20)$~mJy at 1.66~GHz.
In combination with the core flux densities
  obtained at
  5~GHz and the lower flux density limits given by the peak brightnesses of B02, this confirms a flat spectral shape (within the
scatter due to different observations and epochs) at GHz-frequencies, which is shown as a dotted line in Fig.~\ref{fig_ghzflux}.
These findings indicate that the continuous flattening of the single-dish spectrum is evoked by superposition of spectrally steep components, still unresolved with local interferometry but fully resolved and/or separable at VLBI resolution, and by the flat VLBI core, which is still nearly unresolved at the observing frequencies of 1.66~GHz and 2.3~GHz. 
K\"onigl
  (1981) calculated such flat
  GHz-spectra of compact synchrotron sources, involving optical thickness for the synchrotron radiation due to
  synchrotron self-absorption. 
 Thus most of the central radio emission of
  \object{J1101+7225} at these frequencies is emitted by a compact, unresolved
  core.

A significant fraction (up to 50\%) of the single-dish flux density
of the whole galaxy is radiated by the inmost region, still unresolved by
interferometric observations. {\em Relativistic beaming models} explain the extraordinary luminosity of
radio-loud cores with unresolved radio-jet components approaching nearly along
the line of sight at relativistic velocities (e.g. Blandford \& K\"onigl
1979, and see Sect.~\ref{sec_indirect}).

One consequence of these models is a straightforward
explanation of the often observed variability of the core flux density by small variations in the
jet orientation with respect to the observer. This variability is still
observable with single-dish observations in case of core-dominated GHz emission.
Table~\ref{tab_fluxvar} presents single-dish radio flux densities,
measured at 5~GHz at different epochs. The values show a flux density variation up to
20\% between 1986 and 1987. This variability has to be taken into
account and leads to significant changes of short-range\footnote{Due to the power-law relation between flux density and
frequency, the influence of the flux density variability on the spectral index
decreases with increasing frequency interval.} spectral indices in the GHz
domain from different observing epochs. The underlying variability of the VLBI core flux density was  observed at 5~GHz (Table~\ref{tab_ghzspec}) and shows the same minimum at around 1992. Therefore only the later measurements were taken in Fig.~\ref{fig_ghzflux} to approximate the spectral index of the VLBI core.
The observed flux density
variability of the core component strongly supports the concept of beamed radio emission, which can be confirmed in Sect.~\ref{sec_indirect}.

\subsection{\label{sec_propermot}High apparent superluminal motion
  outside the central 10~mas}

\begin{figure}
\includegraphics[width=\columnwidth]{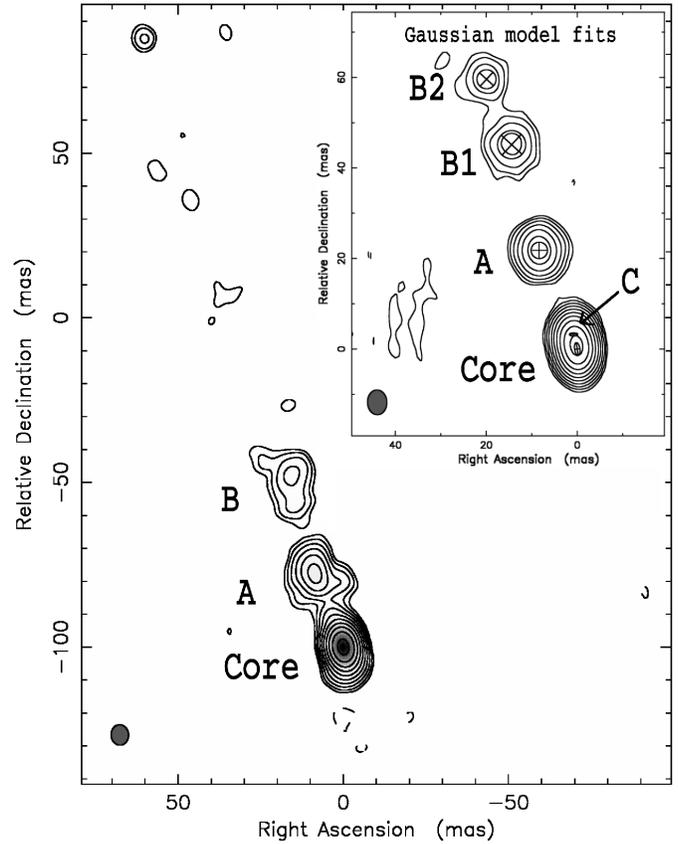}
\caption{\label{fig_EVNmap}Cleaned map of the EVN observation of \object{J1101+7225} at 1.66~GHz. The map shows a peak brightness of
              407~mJy/beam. The contour levels are -0.15, 0.15, 0.3, 0.6, 1.2,
              2.4, 4.8, 9.6, 19.2, 38.4, 76.8\% of the peak brightness, and
              the beam size is 6.22$\times$5.28~mas at position angle (PA) of 4$^\circ$. In the upper right
              corner the Gaussian model fit representation of the VLBI
              image is given (with the same contour levels, but now with respect to a 399~mJy/beam peak brightness), obtained by fitting Gaussian components
              to the visibility data (beam: 5.5$\times$4.3~mas at PA 2$^\circ$). By this technique the extended
              structure at $\sim 50$~mas distance to the core could be
              clearly resolved into two components.
              }
         
\end{figure}

\begin{table}
\begin{center}
\begin{tabular}{cccccc}
\hline \hline
&Flux dens.&\multicolumn{2}{c}{location}&\multicolumn{2}{c}{size}\\
& (mJy) & r (mas) & $\vartheta$ (deg) & a (mas) & Axial
ratio   \\
\hline
Core&  396 &   0  &    0   &       2.88  &  0.41$@$9.4$^{\circ}$ \\ 
C&  149 &   3.3   &   14.1  &   1.83 &   0.39$@$-89.4$^{\circ}$ \\
A&   54 &   23.3  &   21.0  &   3.62 &    1  \\
B1&   16 &   47.4  &   17.7  &   6.22  &   1   \\
B2&    7 &   62.8  &   18.5  &   4.21  &   1   \\
\hline
\end{tabular}

\caption{
\label{tab_gaussmod}The best fit Gaussians as mentioned in the
text. {\em a} labels the major axis of elliptical Gaussians. {\em A}, {\em B1}, and {\em B2} were
only fitted by circular Gaussians, because the  locations of components of lower
flux density can be fitted more reliably with less degrees of
freedom. The flux density errors are $\sim 5\%$, the uncertainties of the
location, and size of the Gaussians range between 0.2-0.5 of the
beamsize depending on the respective flux density.}

\end{center}
\end{table}

\begin{table*}
\begin{center}
\begin{tabular}{c|cc|cc||c}
\hline \hline
 \multicolumn{1}{c}{epoch \& observ. freq. {[}GHz{]}:} & 25 Sep 1991 at 1.66 [T95]& 13 Feb 2002 at 1.66 &\multicolumn{2}{|c||}{mean app. transv. velocity}& 19 Apr 1995 at
 2.3 \\
\hline 
components: & \multicolumn{2}{|c|}{Distances in Fig.~\ref{fig_superlum} $[{\rm mas}]$} &$\mu
 [{\rm mas\,yr^{-1}}]$&$\beta_{{\rm app};h=0.71} \,^{(b)}$ & Dist. in Fig.~\ref{fig_superlum} $[{\rm mas}]$ \\
\emph{Core-A} & $(19.6\pm0.5)$ & $( 23.3\pm 0.5)$ & $(0.36\pm0.07)$& $(22.5\pm4)$& $(21.5\pm3)$  \\
\emph{Core-B} & $(53.7\pm1)$ & $( 52.1 \pm 1 )$ & $0 \,^{(a)}$ & $0$& $(52\pm5)$ \\
\hline
\end{tabular}

\caption{The angular distances between the core and the fitted Gaussian jet components (or their flux density weighted mean; cf. text). The errors reflect uncertainties in the fitting process and increase with decreasing component flux density, while derivation of the apparent velocities is
  described in the text. The position angles of the components did not change significantly during the observing epochs and were omitted. In the last column the read out values from a map by B02 are given for comparison, but cannot be included in the calculations due to lack of a Gaussian model fit. \newline $^{(a)}$: No significant propagation of the
  {\em B} component was found with respect to the errors. \newline $^{(b)}$: For the
  calculations we
    used $H_0=71^{+4}_{-3}\,{\rm km\,s^{-1}}\,{\rm Mpc^{-1}}$ and a deceleration parameter
    $q_0=0.1$.}
\label{tab_compsep}
\end{center}
\end{table*}

\begin{figure}
   \centering
   \includegraphics[width=\columnwidth]{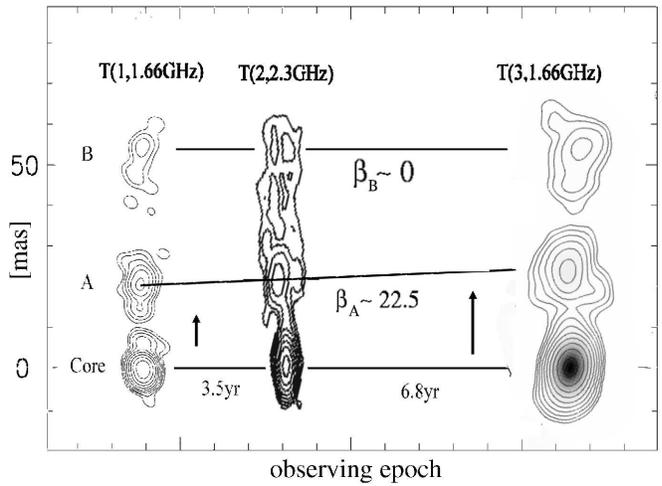}
      \caption{\label{fig_superlum}The cleaned maps are aligned vertically (rotation about -20$^{\circ}$), which is
   justified by the fact, that no significant change of the position angle of
              the jet axis was observed. The apparent linear velocities of the fitted Gaussian components (cf. Table~\ref{tab_compsep}) are
              given. The first map by T95 at T(1,1.66~GHz) shows a peak brightness  of
              476~mJy/beam, the contour levels are -2, -1, 1, 2, 4, 8, 16, 32, 63, 127,
              253~mJy/beam, and the beam size is 4.2$\times$3.2~mas at PA 90$^\circ$. The second map by B02 at
              T(2,2.3~GHz) shows a peak brightness of 520~mJy/beam, the contour
              levels are -3, 3, 6, 12, 24, 48, 96, 192, 384~mJy/beam, and the
              beam size is 3.3$\times$6.7~mas at PA 20$^\circ$. The beam PAs given here refer to the unrotated coordinate system.
              }
         
\end{figure}

In Fig.~\ref{fig_EVNmap} we present the map made from the EVN
    observation in Feb. 2002. Clearly two extended regions are visible beside the core with more than 4.8\% and 1.2\% of the core peak brightness. They are labeled with {\em A} and {\em B},
    respectively. More information about the underlying source
    structure can be retrieved by fitting circular and elliptical
    Gaussians to the visibility data and optimizing the fit to the
    amplitudes and closure phases (cf. Pearson 1995). The structure in 200~mas distance to the core in N-NE direction could be confirmed neither by applying different taper to the uv data nor by a respective model component.
Our best model of the source structure is given in
    Table~\ref{tab_gaussmod} and the corresponding map is inserted in
    Fig.~\ref{fig_EVNmap}. The measurement errors on the model parameters flux density, relative position, and source size are given in the respective tables. Their estimation is based on the comparison of different model fits, as described by G{\'o}mez \& Marscher (2000). A more theoretical approach (Fomalont 1989) based on the dynamic range of the map gives comparable magnitudes. 

This analysis shows that the extended {\em B} component from the
cleaned map is a blend of
at least two components ({\em B1}, {\em B2}). Furthermore within the
central 5~mas a significant fraction of the flux density is
radiated by an extended source beside the
dominating unresolved core at a distance of a few mas to the
core (see Sect.~\ref{sec_centralstruc}). This region close to the limit of
resolution of the 1.66~GHz observations is studied in more detail in the next
section (\ref{sec_centralstruc}) at higher frequencies.

T95 detected both {\em A} and {\em B}, too. A detailed comparison of our data with the Gaussian component model of T95 reveals a high separation
    velocity of the {\em A}-component at an unusually large deprojected
    distance to the core.
In contrast to
    Table~\ref{tab_gaussmod}, they could fit three elliptical Gaussian components to the region around
    {\em A} and one to the {\em B} region. 
Because the angular extensions of both regions ({\em A} and {\em B})
    are close to the limit of resolution, we prefer the {\em blend}
    interpretation of this situation. Our {\em A} component does not correspond to {\em one} of the three fitted components of T95 but does correspond to a flux density-weighted mean of {\em all} three components; and vice versa, the mean of our {\em B1} and {\em B2} components corresponds to the one component of that region fitted by T95 to their visibility data. Although the full width at half maximum (FWHM) of the circular Gaussians
    {\em B1} and {\em B2} (cf. Table~\ref{tab_gaussmod}) are not
    overlapping,  the modeling process showed that due to their low flux
    densities, the sizes of
    the {\em B}
    components are insecure up to a factor of two. It is also
    possible that a third component exists, slightly too weak to be
    fitted. Thus the used flux density-weighted mean appears to be the most
    adequate presentation.

We determined, therefore, the apparent motion of the radio jet structures {\em A} and {\em
    B} with respect to
    the core (Table~\ref{tab_compsep}), where in case of several fit components the labels refer to the flux density-weighted
    mean of these. 
 Angular velocities $\mu$ are presented in Table~\ref{tab_compsep} on
 the right side. They are confirmed by the snapshot VLBA map at 2.3~GHz of
    B02 (last column in Table~\ref{tab_compsep}).  We did not directly include these data in the velocity
    calculation. The introduced uncertainties due to the different observing
    frequency and the lack of a Gaussian modelfit would annihilate the
    increased calculation accuracy.  To demonstrate the apparent angular motion, we show the three VLBI maps in one figure in
    Fig.~\ref{fig_superlum} on a common angular
    scale and rotated by about $-20~^\circ$.

The dimensionless linear
    equivalents $\beta_{\rm app}$ are calculated via $$\beta _{\rm app}=\mu
    \frac{z}{H_{0}(1+z)}\left[
    \frac{1+\sqrt{1+2q_{0}z}+z}{1+\sqrt{1+2q_{0}z}+q_{0}z}\right]$$
    (cf. Pearson \& Zensus 1987). 
The measured apparent superluminal {\em transverse} motion can be transformed
    via special relativity\footnote{\label{foot_beta}If $\vartheta$ is the angle enclosed by the
    line of sight and the direction of motion and if $\beta$ is the intrinsic
    velocity, one finds
    $\beta_{\rm app}=\frac{\beta\,\sin\vartheta}{1-\beta\,\cos\vartheta}$. The minimal
    intrinsic velocity
    $\beta_{\rm min}=\sqrt{(\beta_{\rm app}^2)/(1+\beta_{\rm app}^2)}$ implies an
    angle $\vartheta_{\rm min}$ fulfilling: $\cot(\vartheta_{\rm min})=\beta_{\rm app}$.} into a minimal intrinsic velocity, expressed
    as:
\begin{equation} \label{equ_superlum}
\beta_{min}^A\approx0.9990_{-0.0005}^{+0.0003}\enspace at\enspace\vartheta^A_{min}\approx2.54^{\circ}\,^{+0.55}_{-0.38}
\end{equation}
The $22\,{\rm mas}$ distance to the core at $\vartheta =2.5^{\circ}$ can be deprojected to a large linear distance  of about $4\,{\rm kpc\,(at\,}z=\,1.46,\,q_0=0.1)$.
It turns out that the uncertainties in the estimation of $\beta_{\rm app}$ and
    the Hubble parameter are affecting $\beta^A_{\rm min}$ by less than
    0.1\%. The more distant {\em B}-component was found to be stationary
    within the errors at a mean angular distance of 52.9~mas.

\subsection{\label{sec_centralstruc}Resolved structure in the central 10~mas}
\begin{figure}
\includegraphics[width=\columnwidth]{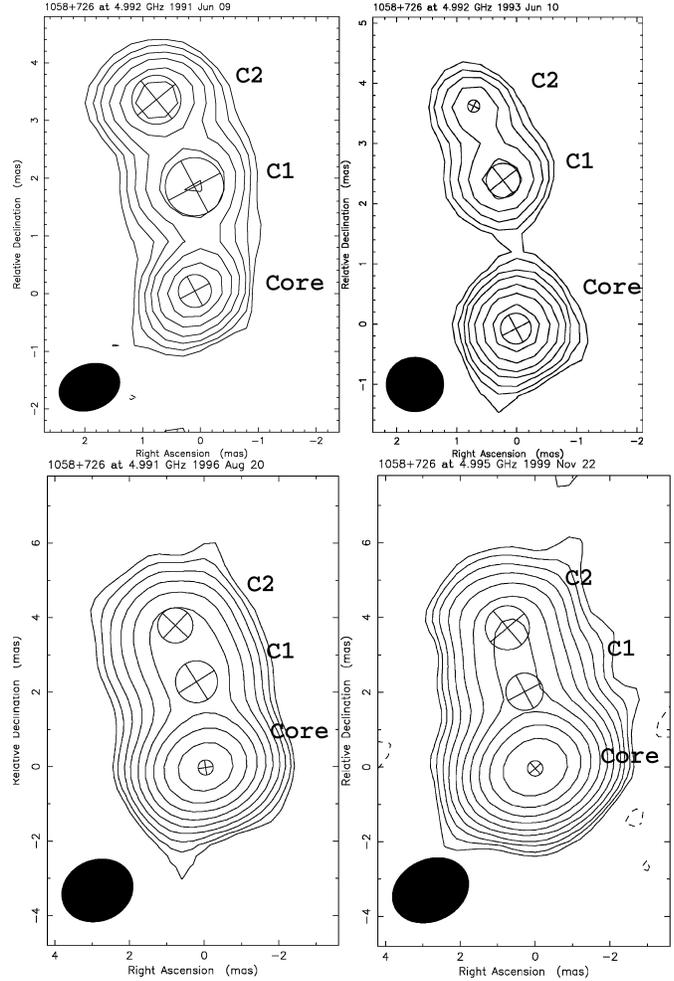}
\caption{\label{fig_GHzsnapshot}5~GHz maps of the central 5~mas. The
  details are given in the form: peak brightness; contour levels in \%; beam
  size. First epoch: 0.097~mJy/beam; levels are -8, 8, 12, 18, 27, 40.5,
  60.8 \%; beam: 1.08$\times$0.79~mas at PA -70$^\circ$. Second epoch:
  0.209~mJy/beam; levels are 6, 9, 13.5, 20.3, 30.4, 45.6, 68.3 \%; beam: 0.96$\times$0.90~mas at PA -90$^\circ$. Third epoch:
  0.314~mJy/beam; levels are 1.6, 2.56, 4.1, 6.55, 10.5, 16.8, 26.8, 42.9,
  68.7, \%; beam: 1.95$\times$1.64~mas at PA -67$^\circ$. Fourth epoch: 0.417~mJy/beam;
  levels are -0.9, 0.9, 1.44, 2.3, 3.69, 5.9, 9.44,
 15.1,  24.2, 38.7, 61.8 \%; beam: 2.11$\times$1.64~mas at PA -65$^\circ$.
Crosses indicate the positions of circular Gaussian
  modelfit components. The detailed model parameters are presented in Table~\ref{tab_britzfits}.}
\end{figure}

\begin{table}
\begin{center}
\begin{tabular}{lccccc}
\hline \hline
epoch&&Flux density &\multicolumn{2}{c}{location}&size\\
&& (mJy) & r (mas) & $\vartheta$ (deg) & diam (mas) \\
\hline
1991.4&Core&  139 &   0  &    0   &     0.57   \\ 
        &C1&  132 &   1.81   &   0.2  &   1.01  \\
       & C2&   133 &   3.38 &   11.3  &   0.85  \\
\hline
1993.4&Core&  282 &   0  &    0   &     0.51   \\
        &C1&  113 &   2.47   &   4.9  &   0.57  \\
         &C2&   61 &   3.76 &   10.7  &   0.20  \\
\hline
1996.6&Core&  337 &   0  &    0   &     0.40  \\ 
         &C1&  97 &   2.32   &   6.0  &   1.13  \\
        &C2&   80 &   3.9 &   12.0  &   0.94  \\
\hline
1999.9&Core&  439 &   0  &    0   &     0.41   \\ 
      &C1&  78 &   2.07   &   7.8  &   1.00 \\
       &C2&   84 &   3.83 &   11.1  &   1.18  \\
\hline
\end{tabular}

\caption{
\label{tab_britzfits}The best fit circular Gaussians of the 5~GHz data over four epochs. The flux density errors are $\sim 5\%$, and the uncertainties of the
location and size of the Gaussians range between 0.2-0.5 of the
beam size depending on the respective flux density.}
\end{center}
\end{table}

\begin{figure}
\includegraphics[width=\columnwidth]{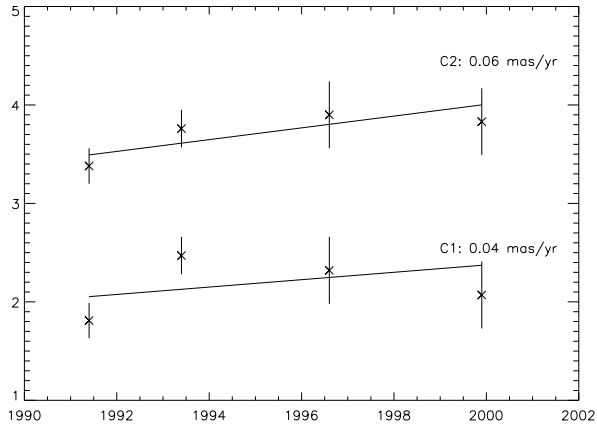}
\caption{\label{fig_regress} Evolving core-distance (in mas) of the best fit Gaussians, taken from Table~\ref{tab_britzfits} over the years. A least-squares fit of a linear time-dependence is shown with respect to the errorbars. The standard deviation of the resulting apparent separation velocities is $\sigma \approx 0.04$mas~yr$^{-1}$.}
\end{figure}

The higher-resolution maps of the 5~GHz snapshots of Britzen et al. (in prep.)
are presented in Fig.~\ref{fig_GHzsnapshot}. They explicitly show that the
{\em Core} region of the 1.66~GHz map is not totally compact on the 1~mas scale of the
5~GHz observations. Extended Gaussian components could be
fitted to the data beside the dominant\footnote{e.g. at the last epoch the
  extended components show flux densities of 20\% of the central flux
  density} central source. This was already suggested by the C-component, fitted to the 1.66~GHz data (Table~\ref{tab_gaussmod}), and it confirms the power of the model fits in analysing the data. 

In Table~\ref{tab_britzfits} the fitted Gaussian components are shown.
A linear fit of the core-distance with respect to the observing epochs (Fig.~\ref{fig_regress}) demonstrates the apparently superluminal motion of both components and shows mean separation velocities of 
\begin{equation}
\, \beta_{app}^{C1}\approx (2.5\pm2.5) \quad \rm{and}\quad \beta_{app}^{C2}\approx(3.7\pm2.5)
\end{equation}

The given standard deviations of the fitted velocities are relatively large due to the few available data points.
Because a significant intrinsic acceleration of the jet material is very unlikely from the
4~mas regime of the inner {\em C}-components toward the 20~mas regime far away from
the central engine, we adopt the $\beta_{\rm min}^A\approx0.999$
from Eq.~\ref{equ_superlum} as the intrinsic separation velocity also for the
inner jet components. From the observed apparent separation velocities of
{\em C1} and {\em C2} one uses the equations of special relativity$^{\,\ref{foot_beta}}$ to calculate {\em two} possible jet orientation angles $\vartheta_{s,l}$ to the line-of-sight for each component (with
$\beta_{\rm min}^A=0.999^{+0.0003}_{-0.0005}$):
\begin{eqnarray} 
\beta_{app}^{C1}\approx (2.5\pm2.5): & \vartheta_s^{C1} \in [0^\circ,0.45^\circ],& \vartheta_l^{C1} \in [22^\circ,180^\circ] \nonumber \\
\beta_{app}^{C2}\approx (3.7\pm2.5): & \vartheta_s^{C2} \in [0.06^\circ,0.56^\circ],& \vartheta_l^{C2} \in [18^\circ,80^\circ]\label{equ_twoangles}
\end{eqnarray}
\noindent
The uncertainty of $\beta_{\rm app}^{C1,2}$ dominates the errors of Eq.~\ref{equ_twoangles}. For both components the calculated orientations differ significantly from $\vartheta^A_{\rm min}\approx(2.54^{+0.55}_{-0.38})^\circ$ of
Eq.~\ref{equ_superlum}.

A closer look at the modelfits in Table~\ref{tab_britzfits} reveals some trends over the observing epochs. The separation motion seems to decrease at later epochs in correlation with a decrease in the respective component brightnesses and an increasing position angle of {\em C1} from 0.2$^{\circ}$ to 7.8$^{\circ}$ over the four epochs. With the given, relatively poor time-resolution of four observations within eight years it is not possible to fit these trends reliably. Nevertheless in combination with the different calculated angles in Eq.~\ref{equ_superlum}\&\ref{equ_twoangles}, they can be interpreted as indications of a spatially curved jet structure. Different line-of-sight orientations account for different brightnesses, position angles, and apparent separation velocities. 

The core flux density has strongly increased over the four epochs, which suggests that over the later epochs either a new jet component has emerged from the compact core region or the unresolved jet has changed its orientation with respect to the line of sight. During the fitting process we found indications that indeed a third component in sub-mas distance to the core is hidden; but 
probably due to its proximity to the core, this new component could not be fitted separately without doubts. 

On the other hand, the increased brightness of the core at the later epochs may affect uncertainties in the model fitting process of the close {\em C} components. The values of one fifth of the beam size (cf. caption of Table~\ref{tab_britzfits} and discussion in Sect.~\ref{sec_propermot}), which are used for the linear fit in Fig.~\ref{fig_regress}, may still slightly underestimate the real errors due to increased overblending of the {\em C} components by the close core radiation.  Because this effect is not quantifiable with our data, we kept using these errors. 

The idea of a jet curvature over the observed core-distances does not rule out a ballistic situation
at the origin of the jet, as was described recently by Stirling et
al. (2003). As origin of the radio jet in BL Lacertae they found  a 'precessing
nozzle' which ejects the single components along straight trajectories. But
outside 2~mas (corresponding to 0.4~mas at the redshift of
\object{J1101+7225}) these trajectories
became curved as well.

Summarizing the motion of all detected jet components including {\em A}
    and {\em B}, the situation resembles
    the different
    measured apparent component speeds of the radio jet of the S5
    quasar \object{0836+710} which
    extends over more than 150~mas at $z\sim2.17$ (Hummel et al. 1992) and where no systematic correlation between the
    component speed and its distance to the core seems to be present (Otterbein
    et al. 1998 and references therein).


\subsection{\label{sec_indirect}Indirect estimation of relativistic bulk motion in the core}
\begin{table}
\begin{center}
\begin{tabular}{cc|c}
\hline\hline
radio & X-ray & results \\
\hline 
\( S_{\rm m}@\nu _{\rm m};\,\nu_2 \) & $\alpha_{\rm X}^{(1{\rm keV};5{\rm keV})}\,^{a)}$  & Doppler factor  \\
 300$\pm 15$mJy@5~GHz;\,2~THz  &  $-0.5\pm 0.2$  & \( \delta =0.5_{-0.3}^{+1.8} \)\\
\hline 
   \( \theta ^{\rm circ}_{\rm m} \)&  \(
S_{\rm obs}^{\rm X}(1{\rm keV})\, ^{a)} \) &  magnetic field \( B_{0} \)\\
\( (0.9\pm0.45)~{\rm mas} \)  &  \( 0.1{\rm \mu Jy} \)  &  $ (0.15^{+0.2}_{-0.12})$G\\
\hline
\end{tabular}

\caption{\label{tab_SSCvalues}The radio values characterize the compact core at 5~GHz as discussed in Sect.~\ref{sec_indirect}. The turn-over properties are mean values of the model fits in Table~\ref{tab_britzfits}. The cut-off frequency $\nu_2$ of the synchrotron spectrum is extrapolated from the optical-radio spectrum. The right column gives the calculated values. 
 $^{a)}$ From
Fiore et al. (2001; 5~keV) and Brinkmann et al. (1997; 1~keV).}
\end{center}
\end{table}

Our VLBI-data of \object{J1101+7225} allow, in combination with published
data at other wavelengths, a straightforward analysis of the
observed VLBI core flux density by applying a {\em relativistic beaming} model.  
The
crucial idea of such models is that due to {\em relativistic} bulk motion of
the sources, the radiation is amplified  or attenuated significantly in the rest-frame of the
observer, depending on the direction of motion with respect to the line of sight. 
Of course the bulk motion
cannot be observed directly due to the lack of spatial resolution. 
But we share the
common assumption that the X-ray emission of the core consists mainly of synchrotron
photons scattered to shorter wavelengths via the inverse Compton
effect. Marscher (1983) quantitatively derived the connection between radio
and X-ray flux densities, their spectral indices, and the Doppler factor\footnote{With the terms of
  footnote~\ref{foot_beta} it is $\delta =
  (\gamma\,(1-\beta\,\cos\vartheta))^{-1}$ with $\gamma=1/\sqrt{1-\beta^2}$} $\delta$ for a homogeneous
source of spherical shape. This implies that $\delta$ can be estimated from
the other {\em measured} values.  

In Table~\ref{tab_SSCvalues} the ingoing values of the calculations are given.
The notation of the radio properties follows Marscher~(1983). Typically the unresolved core consists of several components each of which dominates the unresolved spectrum at a different frequency (e.g. Marscher~1988). Therefore we can assume that the measured properties at 5~GHz describe the {\em turn-over point} (index {\em m}) of one underlying source component and that the formulae can be applied for that component. The X-ray measurements were taken independently in the 1990s. Therefore a mean of the different 5~GHz measurements is the most appropriate way to combine both datasets.

The mean measured Gaussian source size is transformed into $\theta^{\rm circ}$ assuming an underlying circular source shape (see Marscher~1983 and Table~\ref{tab_SSCvalues}). The angular source size  should be understood as an order of magnitude
estimate\footnote{If much more (spectral and multi-epoch) data are available
  as e.g. in the case of the quasar \object{3C 345}, inhomogeneities in the particle
  number density can also be estimated. Lobanov \& Zensus (1999) used more
  detailed analyzis to explain the observed flux density
variations quantitatively with time of \object{3C 345}.}. Further we can use only the measured peak flux density $S_{\rm m}$ in contrast to the extrapolated one as foreseen by Marscher~(1983). This results in a slight underestimation of $\delta$.
The calculations lead to $\delta =(0.5_{-0.3}^{+1.8})$ and  a magnetic field
strength of $B_0$=$(0.15^{+0.2}_{-0.12})$~G of the unresolved
synchrotron-self absorbed source (cf. Table~\ref{tab_SSCvalues}). 

However further radio components, peaking at higher frequencies, and other processes than inverse Compton scattering may contribute to the observed X-ray flux densities. This results in a probable underestimation of  $\delta$ and $B_0$. In fact a flat to inverted spectral index as indicator for several components is suggested by the VCS1 map at 8.4~GHz (B02).
Although the lack of a detailed model fit inhibits reproducible calculations, visual inspection of the map reveals a dominating core, probably unresolved. Under the latter assumption the given peak brightness resembles the flux density of the component in mid 1995, which is slightly larger than the respective values of the 5~GHz experiment at that time (cf. Fig.~\ref{fig_ghzflux}).

With the angular source size and the Doppler factor, the intrinsic brightness temperature can be calculated as 
\begin{equation}
T_{intr} = T_{app}/\delta = (2.8^{+0.5}_{-0.2}\cdot10^{11}\,{\rm K})\quad.
\end{equation}
These values fit well in the range of brightness temperature
in which the simultaneously observed synchrotron emission and the inverse
Compton-scattering are both effective enough to fulfill the {\em inverse
  Compton} scenario adopted here ($2\cdot 10^{11}~{\rm K} \leq T_{\rm IC} \leq 10^{12}~{\rm K}$; e.g. Kraus~1986, Bloom \& Marscher~1991). Without applying a Doppler factor the {\em apparent} brightness temperature at 5~GHz rises from 0.5-3.1$\cdot10^{11}~{\rm K}$ over the four epochs. Both this strong rise and the fact that the earlier estimates are below the lower inverse Compton limit can be interpreted as indicators of relativistic beaming.

Furthermore we calculated the equipartition Doppler factor
  $\delta_{\rm equ}$ following Readhead
  (1994) and G\"uijosa \& Daly (1996), which implies that the radiating
  particles have the same energy as the penetrating magnetic field. The
  results (assuming as above $h=0.71$) 
\begin{eqnarray}
\delta_{equ}&=& (0.5^{+2.6}_{-0.3})
\end{eqnarray}
are very similar to the inverse-Compton Doppler factor (Table~\ref{tab_SSCvalues}). This supports Readhead's conclusion that many powerful, non-thermal extragalactic radio
sources are close to energy equipartition and their finding that the brightness temperatures of the respective sources range far below the maximum brightness temperature of $10^{12}~{\rm K}$ (see above).

As in the previous section, we now adopt that the intrinsic velocities along
  the jet are equal to the highest minimal intrinsic velocity, as given by the
  apparent separation speeds. Then the angle between the direction of motion and the
  line of sight can be estimated from the Doppler factor:
\begin{equation}\label{equ_dopplerspeed}
\delta_{core}=(0.5_{-0.3}^{+1.8});\beta_{min}^A\approx(0.999^{+0.0003}_{-0.0005})\rightarrow\vartheta_{core}\approx(24^{+19}_{-14})^\circ
\end{equation}
Thus $\vartheta_{\rm core}$ includes only the large angle solutions of Eq.~\ref{equ_twoangles} and supports the idea of a straight inner jet nozzle as described in Stirling et al.~(2003).

\section{\label{sec_sum}Discussion of the results}

For the first time the kinematics of the non-thermal radio jet components
of \object{J1101+7225} were estimated.
We calculated the apparent motions of the two
extended components, observable at low GHz-frequencies from VLBI-measurements. We found an apparent superluminal separation speed of \( \beta_{app}(Core-A)=(22.5\pm 4) \)
for the \( A \)-component over the past decade at an exceptionally large deprojected
distance to the core $(22\,{\rm mas}\sim\,4~{\rm kpc})$. 
The even more distant \( B \)-component
was found to be stationary within the errors at a core-distance of $\sim 53~{\rm mas}$. 

Furthermore a Doppler factor $\delta=(0.5^{+1.8}_{-0.3})$ was estimated for the compact
optically thick (at the observing frequencies) synchrotron radiation
of the core of
\object{J1101+7225}. The strength of the magnetic field, which is necessary for the synchrotron
process, was estimated to be $B_0=(0.15^{+0.2}_{-0.12})$~G. The estimated size of the optically thick nuclear component is close to the resolution limit of an interferometric observation at
cm-wavelengths, which suggests, in combination with the estimated
Doppler factor and intrinsic jet velocities, that additional VLBI-components could appear
outside the core in the future. 

More observations of these nuclear components could reveal, if a
deceleration of the \( A \)-component appears while {\em A} is approaching
the  \( B \)-component. This would be similar to the well known case of
\object{4C 39.25} (Alberdi et al. 1993a).
  Deceleration at
these distances from the core can confirm an interaction of the jet
component with the circumnuclear matter of the host galaxy as described by
Taylor et al. (1989) for Seyfert nuclei. Such an interaction is strongly
  supported by comparing the VLBI maps presented  here with the jet
  geometry as observed with the VLA (Xu et al. 1995). In their 1.4~GHz map two
  corresponding radio jets appear, extending along $\sim 5''$ from the
  core. While the southwestern jet extends in the opposite direction of
  the VLBI jet, the northeastern VLA-jet does {\em not} coincide with the position angle of the VLBI
  structure. Instead it appears to be bent towards a northwesternly direction,
  perhaps induced by ram pressure of the surrounding material. 

{\em Apparent} superluminal velocities are explained by
motion towards the observer at relativistic velocity. The additionally
  presented 5~GHz maps and Gaussian modelfits of the
central 5~mas show further jet components with significantly smaller apparent velocities. Thus the
separation velocities of the different VLBI radio jet components of
\object{J1101+7225} show no simple correlation with their
distance to the core. But this does
{\em not} rule out a constant intrinsic velocity, because already small
variations of the angle between jet and the line of sight can introduce
variations of the apparent speeds of the observed order of magnitude. In the
case of such a constant intrinsic velocity along the whole jet, the very
    high velocity, derived from the \( A \)-component, has to be chosen. Such
    high intrinsic velocities explain both the large extension of the radio jet over several arcsec as
    estimated with the VLA and the high luminosity of the core.

We
believe that our findings may indicate a helical
bending of the jet, where the fast components are moving in a section
that is curved towards the observer. This is supported by the estimated differing jet orientations with respect to the line of sight.
Zensus et al. (1995) could explain the acceleration of a jet component of
\object{3C~345} by a curved jet of constant intrinsic bulk velocity. It has been shown
  that helical jet patterns result from Kelvin-Helmholtz and
  current-driven jet instabilities in relativistic flows (Birkinshaw 1991,
  Istomin \& Pariev 1996).

Beside this interpretation of the observed motion as an intrinsic bulk motion
of the radiating plasma, shock waves may travel along
the jet (Alberdi et al. 1993b).
Different component velocities may also be observed if the slower components
are {\em trailing} in the wake of faster ones. This hydrodynamical explanation
could be adopted successfully to the complex component motion in the radio jet of
the radio galaxy \object{3C 120} (Gomez et al. 2001). Hardee et al. (2001)
  found mechanisms to produce differentially moving and stationary
  features in a jet by analyzing the relativistic hydrodynamic equations.

Thus the nuclear region of \object{J1101+7225}
provides the rare possibility of observing the total range of jet kinematics
including apparent superluminal separation velocities
even far out of the central parsec-region. The results presented here can
  give observational constraints far from the jet origin for
numerical jet models.

\begin{acknowledgements}
We are grateful to the correlator team of the MPIfR in
Bonn for their assistance. 
The European VLBI Network is a joint facility of European, Chinese, South African, and other radio astronomy institutes funded by their national research councils. The National Radio Astronomy Observatory is a facility of the National Science Foundation operated under cooperative agreement by Associated Universities, Inc.
This work was supported in part by the Deutsche
Forschungsgemeinschaft (DFG) via grant SFB 494. 

\end{acknowledgements}

\end{document}